# Bitcoin Cooperative Proof-of-Stake

Stephen L. Reed

**ABSTRACT**

A hard-fork reconfiguration of the peer to peer Bitcoin network is described that substitutes tamper-evident logs and proof-of-stake consensus for proof-of-work consensus. The block creation rewards and transaction fees are reallocated to establish and staff a secure financial data network capable of handling the world's transactions with subsecond response time. The new system pays dividends to stake-offering bitcoin holders. In contrast to Satoshi Nakamoto's mesh network consisting of competing peers, this system uses an enterprise class network that is efficient, robust, and scalable, consisting of cooperating peers. The network backbone nodes host trustless nomadic agents. Thousands of distributed full nodes are paid to replicate a singleton blockchain built upon every 10 minutes by a nomadic mint agent whose actions are verified by its peers. This arrangement enables immediate acknowledgment to an issuing node that its transaction has been accepted. Less effort means that subsidized transaction costs will be lower. Network reconfiguration enables the processing of numerous microtransactions. Stake-weighted distributed consensus is achieved when necessary with less than one-half arbitrarily faulty nodes. Important invariants of the Satoshi Social Contract between core developers and users are maintained: The reward schedule, the blockchain format, the fixed number of bitcoins, and the decentralized, trustless protocol are untouched. The system remains a global distributed database, with additions to the database by consent of the majority, based on a set of transparent rules they follow.

**1. Satoshi Nakamoto's Bitcoin and the Current Situation.**

Satoshi Nakamoto's [1] Bitcoin software assumes an arbitrarily faulty, insecure, and adversarial peer to peer environment. No distinction is made with regard to the capabilities of full node peers, nor of the mesh network connections between them. Accordingly, no provision is made for efficiencies gained via specialized peers, no provision is made for cooperating peers, nor is there any provision or incentive for network infrastructure, for security or for operations management.

Transactions are relayed and accepted into the blockchain on a best effort basis. A new transaction reaches 50% of nodes within 1.2 seconds and 90% of nodes within 2.9 seconds [2]. Transactions are not timestamped. There is no definite version of the blockchain. In contrast, incumbent credit/debit card payment systems are faster [3] and more certain for consumers. Incumbent bank wire transfer, e.g. Swiftnet [4], is faster and more certain for business-to-business users. Incumbent payment transfer systems have data security policies that Bitcoin lacks [5] with regard to protecting host computers and customer data, e.g. private keys.

A few centralized pools, operating ad-hoc infrastructure on thin 2% commissions [6], create most new blocks. Most of the remainder are created by a few competing public cloud and private data centers, including unspecified mining device manufacturers.

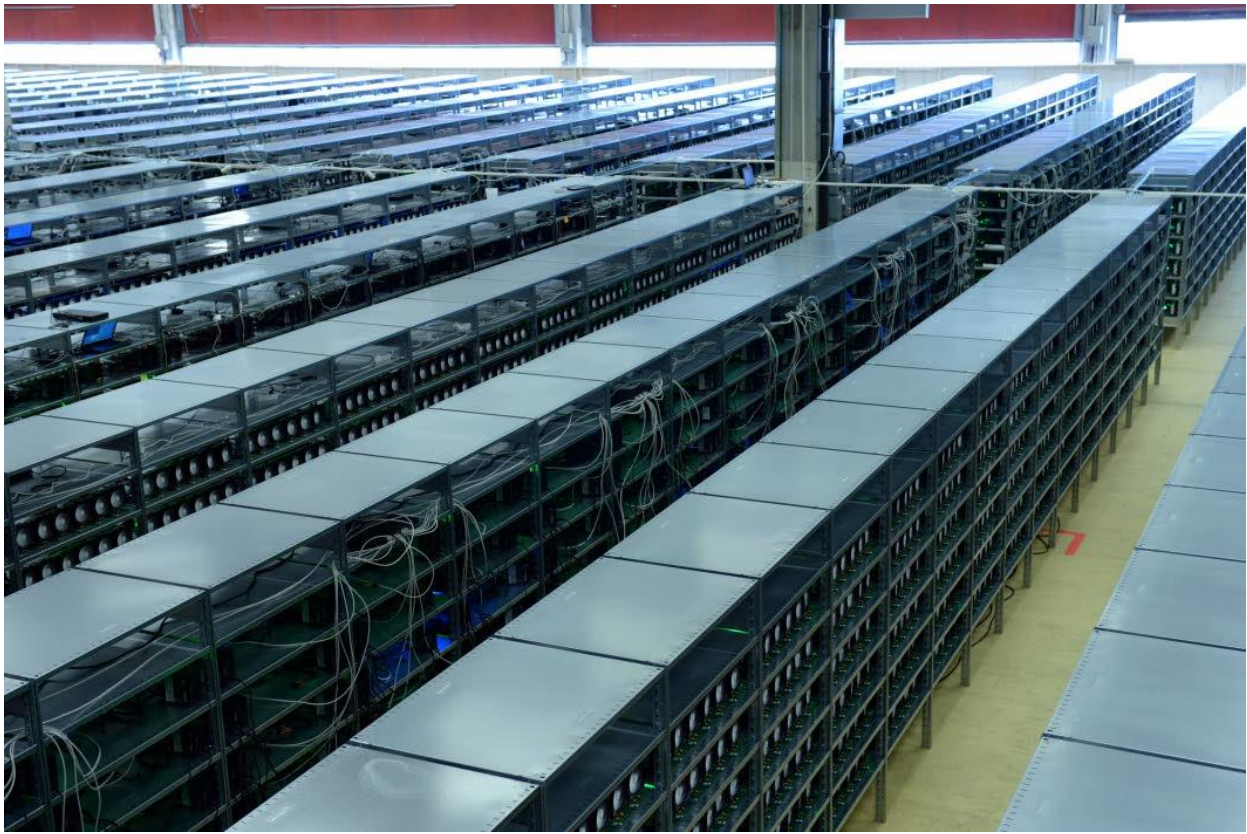

Uncompensated full node owners, burdened by operating costs, are dwindling in number[7] [8], and are geographically concentrated in the Eastern United States.[9]

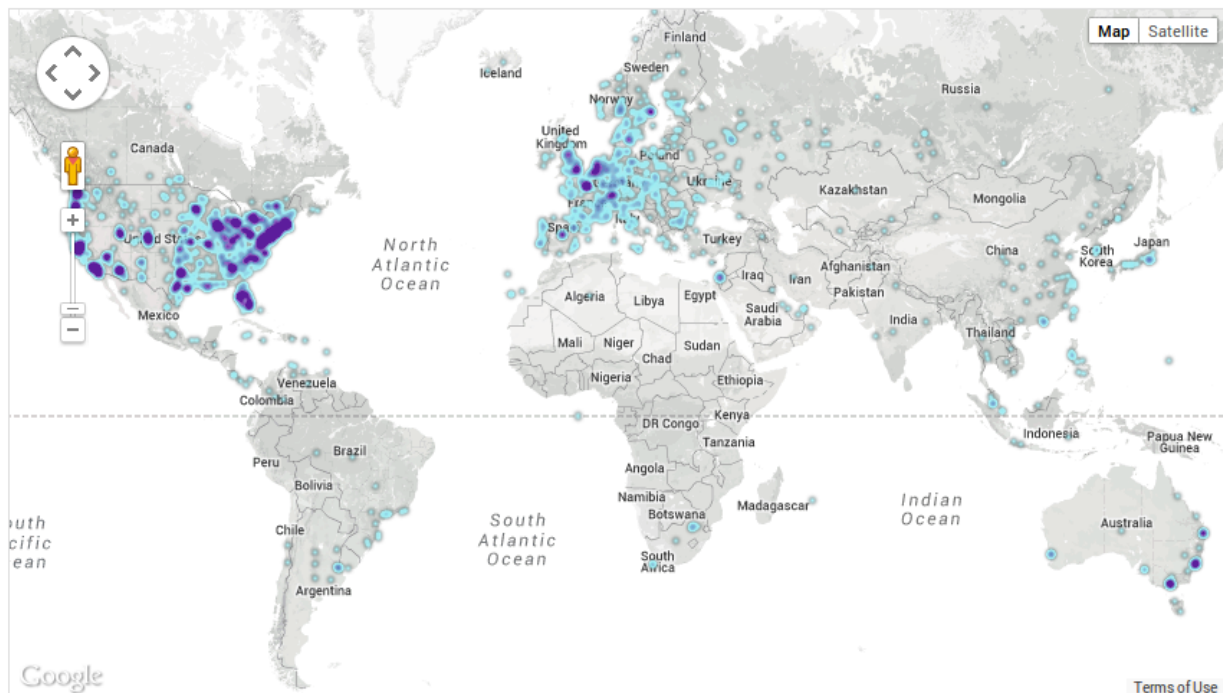

Map shows concentration of reachable Bitcoin nodes found in countries around the world.

Proof-of-work mining by design spends block creation rewards almost entirely on single purpose hashing devices and power. In the past year, that amount totaled $490,103,250. The following logarithmic chart illustrates the rapid growth of miner's daily USD revenue over the last two years [10].

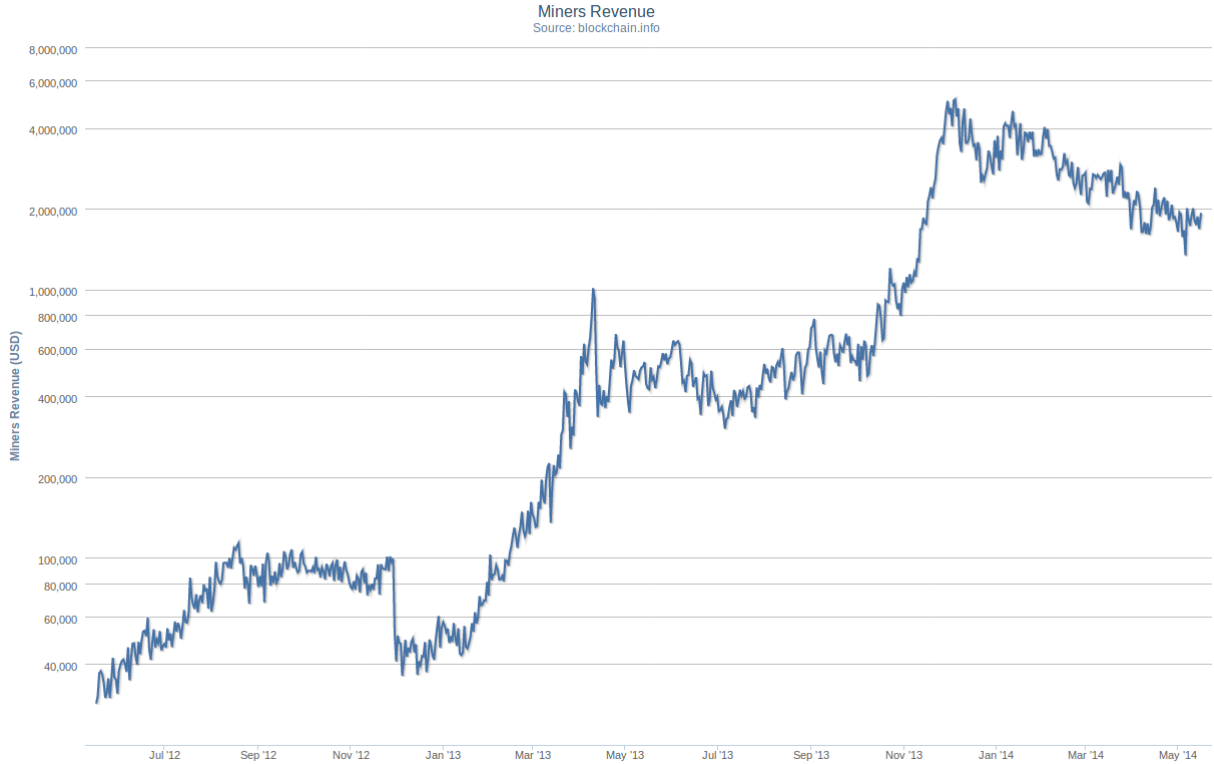

Understandably, Satoshi Nakomoto's Bitcoin has been dubbed a "peer-to-peer heat engine" [11].

## 2. A Simplified Model of Satoshi Nakamoto's Bitcoin.

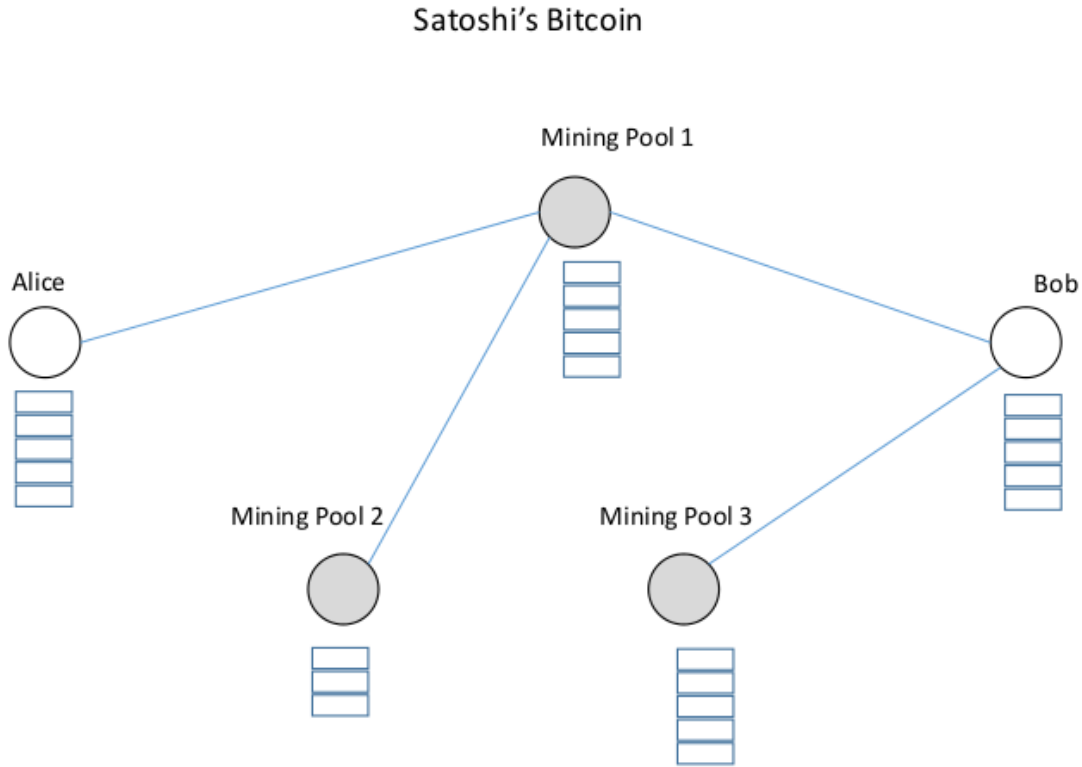

In the simplified figure above, Alice pays Bob with a new payment transaction issued into the randomly connected network [12][13][14][15]. Alice's transaction is verified and relayed by each receiving node to respective peers who do not yet have the transaction. Mining pool nodes choose whether to accept Alice's transaction into a new block, and compete with other mining pools to create a hash proof-of-work meeting the current network difficulty. The winning mining pool pays its participating hashers with the block creation reward and included transaction fees. The reward may not be spent until after 100 additional blocks have been created. Alice expects, but is not guaranteed, that her transaction will be included into a new block within an average of 3 minutes. The new block she receives has a certain low probability of reversion, decreasing with time, should another mining pool create a superior blockchain with greater aggregate difficulty. The longest chain rule ensures practical convergence on a consensus blockchain after about 6 new blocks have been added after the block containing Alice's transaction, which is about an hour at the rate of 10 minutes average between mined blocks. Bob inspects his consensus version of the blockchain to ensure that the transaction is confirmed, i.e. accepted into a block.

## 3. Alternative Cryptocurrency Innovations.

Satoshi Nakamoto's solution to the double-spending problem requires the thermodynamic effort of hashing to achieve distributed consensus [16]. The first, yet impractical, improvement was to somehow obtain efficient unbounded agreement, i.e. the consensus that permits blockchain checkpoints in the code, and randomly permit a full node to create the new block without an onerous proof-of-work calculation. [17].

Subsequent innovations to address proof-of-work effort were implemented as alternative cryptocurrencies. PrimeCoin [18] substitutes the purported useful work of finding certain new prime numbers for the Satoshi method of SHA-256 hashing. If Bitcoin used this method, the amount of thermodynamic effort would be reduced a tiny bit by the benefit of finding these numbers.

Proof-of-stake was first mentioned in July, 2011 by QuantumMechanic [19]. Stake-voting as a method to achieve distributed consensus for building the blockchain was not available as an option to Satoshi because Bitcoin started in 2009 with no stake. By July 10, 2011, the market capitalization of bitcoin had risen to $106,345,680. Subsequently, two proof-of-stake algorithms were authored in 2011 and are described in the Bitcoin Wiki [20],

PeerCoin [21] is the cyrptocurrency with the fourth highest market capitalization [22], featuring proof-of-stake as an adjunct to proof-of-work, ultimately obviating the need for a calculation effort. The stake is determined by Coin Age - a metric favoring large long-unspent transaction inputs [23]. Critics point out the peers may contribute stake to more than one competing branch of the blockchain [24][25].

Numerous alternative cyrptocurrencies based on proof-of-stake [25] however retain the original Satoshi design of adversarial peers racing against each other to build new blocks on a branching blockchain. They typically tweak parameters of the original Satoshi implementation to offer faster confirmation times, differing maximum currency limits, and differing block reward schedules.

**4. Cooperation Without Trust.**

Previous research into cooperating trustless agents by Nick Szabo (1997) [27], (1997) [28], and (1998) [29] argued that authority is more trustworthy when distributed, in particular when there is a separation of powers. Szabo proposed a quorum of peers to approve actions by an agent otherwise subject to fraud. He proposed unforgeable auditing logs, secured by one-way hash functions, that auditors could review to ensure the absence of fraud. Andrew Miller et. al. [30] described a logical language for the description of authenticated data structures whose operations are performed by an untrusted agent and which can be verified by another agent. Maniatis and Baker [31], describe the process whereby such logs can be made tamper-evident through timeline entanglement.

According to arguments presented by Szabo [32], trusted third parties introduce security risks into a security protocol and should therefore be minimized in the design. He mentions Byzantine (arbitrary-faulting) resilient replicated databases, e.g. Fleet and Phalanx, which avoid dependence on trusted third parties.

PeerReview [33] is a C++ library which has provides accountability for a variety of distributed systems, providing them with the means to justify their actions to their peers.

**5. A Single Mint Bitcoin System.**

In contrast to Satoshi's Bitcoin, the proposed system exhibits coordinated and cooperative behavior. According to Szabo, trustless cooperation is possible in a distributed system through authenticated peer attestation of correct behavior. Consensus, achieved in this system by stake-weighted voting, is primarily required when misbehavior is detected. The method is resistant to adversaries lacking sufficient stake. A faulty or misbehaving peer is disconnected from the network by a quorum of its stake-weighted peers.

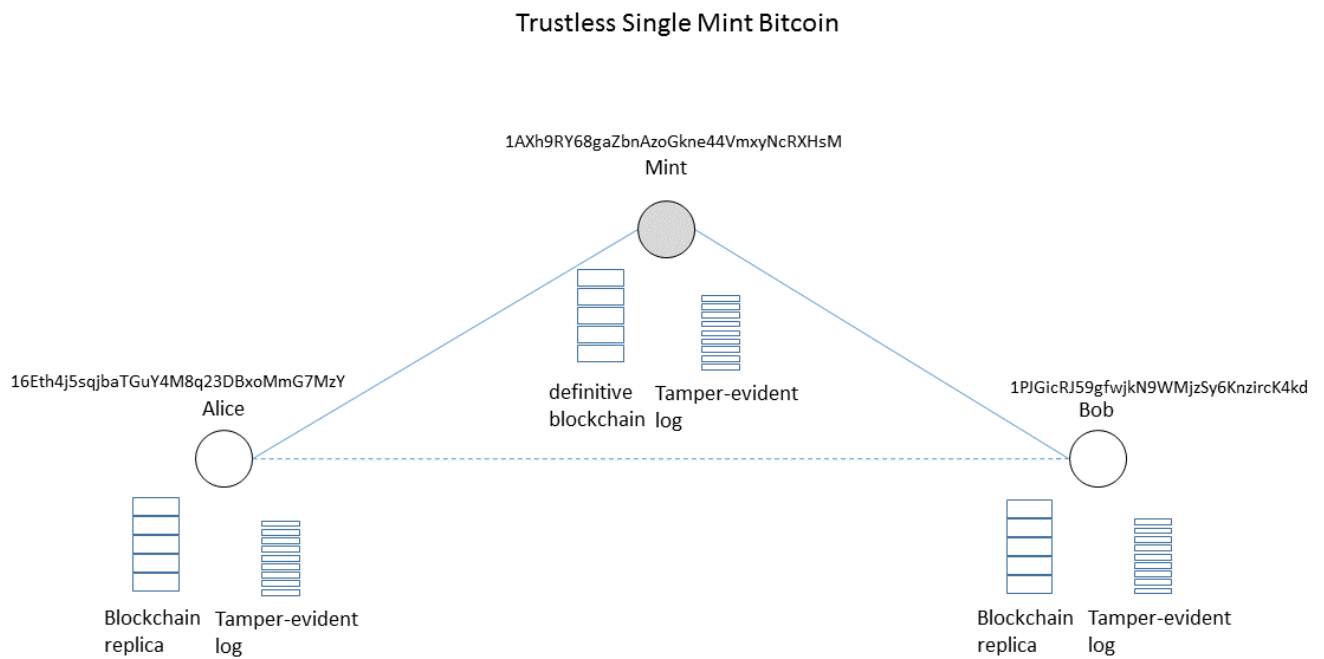

In the simplified figure above, a central trustless mint agent creates all new blocks, appended to a non-forking blockchain, on a fixed schedule, without proof-of-work effort.

Alice prepares to pay Bob with a new payment transaction. Her tamper-evident log records the issuance activity details that include the local timestamp, the Bitcoin transaction, and the connection endpoint's, e.g. the mint's, bitcoin address. The transaction to be sent is packaged with the transaction details, and an authenticated hash of Alice's tamper-evident log. The transaction sent directly to the connected mint.

The mint directly receives the transaction and logs its arrival. The mint immediately sends back to Alice an authenticated hash of its own tamper-evident log containing her transaction. At this point the tamper-evident logs of Alice and the mint are entangled [31]. Despite differences in their respective local clocks, the temporal order of the distributed process steps can be verified by any observer querying the logs. Alice knows that the mint received her transaction.

The mint agent validates Alice's transaction. If invalid, that status is logged and it sent back to Alice. Otherwise, the transaction is a candidate for inclusion into the new block. If there is no transaction fee, the transaction may be accepted according to a rule that permits a certain percentage of free transactions per block. If rejected by that rule, that status is logged and it is sent back to Alice. Otherwise the mint logs the accepted transaction status.

The mint broadcasts the accepted transactions directly to its peers, namely Alice and Bob. Bob immediately knows that Alice's payment to him will be included in the next block created by the mint with trivial proof-of-work at exact 10 minute intervals. When the mint creates the new block, it keeps a certain portion of the block reward and transaction fees for itself, and pays the remainder daily to Alice and Bob as immediately spendable dividends in proportion to their offered stakes. Bob and Alice both log the acknowledged transaction, and verify the work of the mint by building their own new block using the same inputs in the same timestamped order as the mint. Tamper-evident logs from Alice, the mint, and Bob permit each of these participants to prove that they behaved correctly. Each participant identifies itself by a certain bitcoin address that they respectively control. The private key of the address is used to digitally sign certain messages and to verify that responses were sent by the intended recipient.

Because there is a single mint, there is no need for a proof of work. The tamper-evident logs facilitate remote attestation of correct peer behavior. The identity of Alice and Bob are supported by their respective stakes. When there is a need for a consensus, votes are tallied and weighted by offered states. Misbehaving nodes are banned from the network, thus strongly motivating honest behavior.

## 6. The Proposed Decentralized Enterprise-Class Bitcoin System.

As Szabo points out [32], the simple trustless mint system described above has the obvious problem of centralization and lack of redundancy. A fault in the mint causes system failure, and the mint is a certain target for attack. The proposed bitcoin system distinguishes the behavior of the mint agent from its host node. The mint agent is nomadic, and can move its process state from one host node to another on a consensus schedule - or as otherwise needed, e.g. fault recovery.

This system is implemented as a hierarchical network of transparent nomadic software agents executing a common open-source software program with differing roles. There are no humans in the loop except as owners of nodes which host the agents. The human node owner may offer a controlled bitcoin address containing stake in return for daily dividends, but is otherwise anonymous.

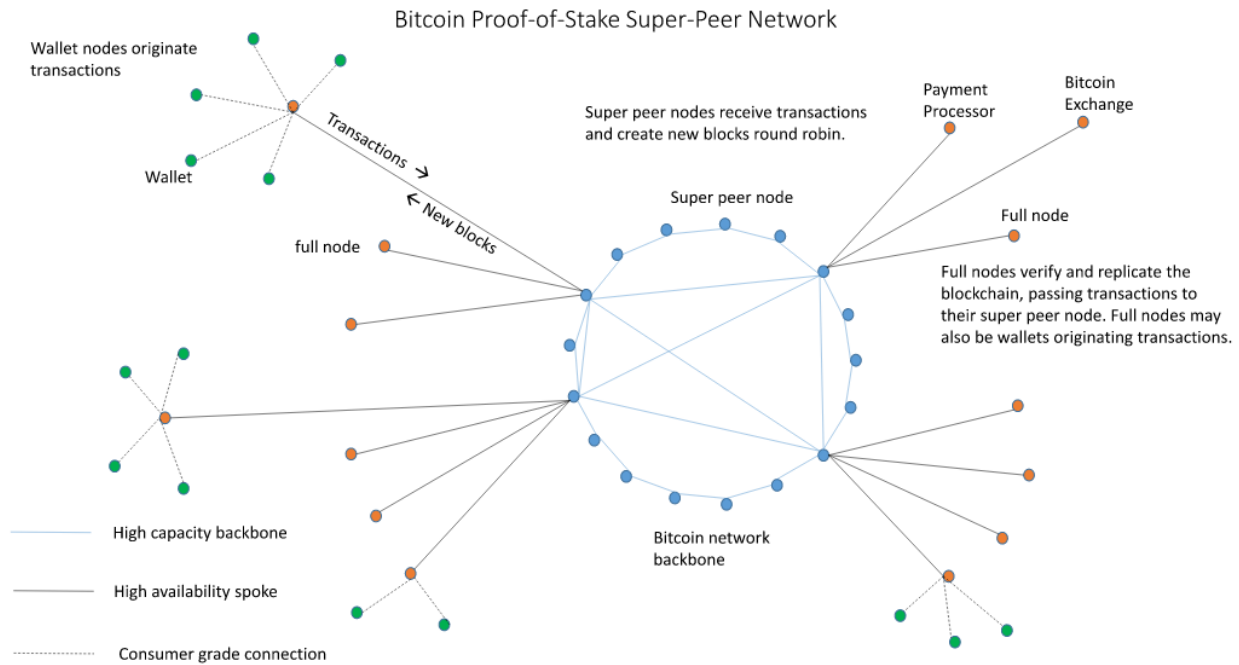

The paid-for enterprise class network capable of handling all the world's transactions is configured as a super peer network [34], in which certain peers have distinguished superior capabilities that allow them to be the network backbone [35]. Bitcoin message traffic originates in the peripheral nodes and flows directly through the backbone to reach the Mint agent. Acknowledged transactions and new blockchain hashes flow outwards from the Mint to the blockchain-replicating full nodes, and to wallets beyond them. Single direct connections are shown in the figure above, however each node has multiple connections to super peers carrying redundant messages, for fault tolerance and misbehavior detection. Each super peer is a hub for numerous full nodes. To service billions of endpoints and numerous microtransactions, the super peer ring may be augmented by one or more outer rings that aggregate inbound message traffic, and conversely disaggregate outbound traffic. Global backbone nodes situated near carrier points-of-presence and the hub-and-spoke network architecture yield sub-second response times given the simplicity of the single append-only journalling Bitcoin database - its blockchain, and append-only logs. Subsecond response time is also enabled by the 6 hop round trip path to the mint. Scaling to additional super peer rings would add only 2 round trip hops per ring.

Super peer nodes are likely to be branded entities such that a few branded entities control the majority of super peers. Important full nodes that serve hosted wallets, payment processors, exchanges, investment firms and financial institutions are likely to be preferentially connected to reliable branded super peers, if not actually colocated physically and operated under the same brand.

This system uses an attestable unforgeable log organization inspired by Nick Szabo. In particular, this system uses attested append-only memory as described by Chun et. al. [36] who provide mathematical arguments for is properties. It remains correct and keeps making progress even when half the replicas, e.g. blockchain replicas, are faulty. This is an improvement over previous Byzantine Fault Tolerant algorithms lacking a tamper-evident log, which allowed only one third faulty replicas [37].

### 6.1. What All Nodes Know.

All full nodes have a built in set of bootstrap node DNS URIs and a built in set of likely good node IP addresses. All full nodes have consensus agreement as to the current set of super peers. All full nodes have a synchronous replica of the single non-forking blockchain and consensus agreement as to the latest block hash. All full nodes have consensus agreement as to the new transactions that have been issued since the last block creation, and have consensus agreement as to the timestamped order of acknowledged transactions. All nodes have consensus agreement as to the approximate date and time, and agreement as to the calendar of super peer agent assignments and activities, e.g. the network reconfiguration each weekend by a designated super peer.

### 6.2. Behavior Common to All Full Nodes and Super Peer Nodes.

Each full and super peer node has identical software, which is capable of performing any particular role, or attesting the role of a peer by replaying that peer's logged operation with logged inputs and outputs.

Each node validates and maintains a copy of the singleton blockchain and the current hash.

Each node owner may be motivated to contribute a test node, according to the reward distribution policy.

Each node controls a Bitcoin address whose private key is used for digital signatures.

Each node maintains a tamper-evident justification log of its operations, inputs, and outputs in JSON format, which is a set of attribute-value pairs which are machine readable by most computer languages, and also by humans. An API permits the log to be read by a peer. The purpose of the log is to attest correct behavior, isolate misbehavior, and to automatically diagnose faults.

Each node maintains connections to a set of random peers whereby a stake-weighted majority can resolve certain disputed state according to a byzantine, i.e. arbitrary, fault tolerance algorithm.

The associated firewall between the node and the Internet may be configured to deny all connections except those in the list of permitted IP addresses. In this configuration, a manual enrollment step is required before the peer node can connect to a given whitelisted peer.

The Bitcoin protocol does not actually require encryption, as the transactions, messages, blocks and the blockchain are cryptographically verifiable. Pseudonymous addresses, dates and transaction inputs/outputs are the only customer data revealed.

**6.3. A New Full Node.**

A new full node joining the network is assigned a set of peers by the user-selected DNS bootstrap node. The new node crawls the graph of connected nodes, constructing a set of random nodes. The new node determines the fitness of its peers [38], which in return determine the new node's fitness. The utility function vector scores uptime, inbound/outbound bandwidth, latency, degree of connection redundancy, CPU throughput, and blockchain presence. It is expected that ordinary full nodes will have sufficient symmetric bandwidth for relaying transactions and maintaining the blockchain. Candidates for super-peer status will be expected to have enterprise class bandwidth, fail-over redundancy, CPU throughput, DDoS protection, etc., insofar as these can be measured by a software probe. The new node solicits the members of the set of super-peers from ten random connected nodes. Using optional manually-configured super-peers as an override, the new node connects to three super-peers having low latency, provided they are eligible to accept additional full node peers. For example, if the Bitcoin super-peer network is configured by core developers at 100 super peer nodes, then each should receive a maximum of 10% of connections - providing a certain amount of decentralization. One of the three connected super-peers is the primary connection to the network, the other two are backups which carry redundant traffic.

The new node provides justification whether it should be a member of the super-peer set, when asked by the unique configuration node.

Each full node provides an API and a digitally-signed machine-readable justification log of its actions, e.g, issued transactions, forwarded transactions, received transactions, new broadcast block hash received, which are verified randomly by the audit node, and by its peer nodes.

**6.4. Bootstrap Node.**

The bootstrap node is a full node having the permanent responsibility of returning a candidate set of connected peer nodes to a requesting new node joining or rejoining the Bitcoin network.

### 6.5. Behavior Common to All Super Peer Nodes.

Super peer nodes may be provisioned as multiple computers in a cluster with load balancing, provided that their temporary responsibility is suitably parallelizable.

Upon the scheduled termination of a super peer node's temporary role, its process state is serialized into a digitally signed JSON data structure and passed in a message to the next super peer node configured to have that role. This behavior allows the software agents to move around the network as nomads. Software code does not move, the process state moves, as all nodes execute identical software.

### 6.6. Configuration Agent.

The singleton configuration agent, hosted by a super peer node, has the responsibility of choosing which nodes are super-peers and setting the nomadic agent schedule. By crawling the network, it records the utility indicators of each node and selects the best N, i.e. 100, nodes as super peers. It instructs them via a message to ensure direct, possibly redundant, connections between them. At the conclusion of the minting cycle, the configuration agent randomly chooses the next configuration agent from among its super-peer nodes.

The configuration agent periodically, e.g. weekly, solicits to-myself proof-of-stake transactions from peers that indicate willingness to offer stakes. These stakes establish the default reward portion.

### 6.7. Seed Agents.

The seed agents, hosted by super peer nodes, have the responsibility for seeding full nodes that join the network. The seeded full node selects the seeding agent having the optimal combination of low latency and current load.

### 6.8. Mint Agent.

The singleton mint agent, hosted by a super peer node, has the responsibility of creating new blocks and minting new bitcoins. New block creation is synchronous, using the consensus timestamp at the hour, 10 minutes past the hour, 20 minutes past the hour, . . . at the rate of six blocks per hour. Received transactions are immediately broadcast back into the network with the acknowledged arrival timestamp, so that all full nodes may build the new block in synchronization with the mint agent. When the issuing full node receives the acknowledged transaction, it may indicate to the user that the transaction is acknowledged, i.e. good-to-go, but not yet committed into the blockchain.

After aggregating new transactions for 10 minutes, the mint agent creates the block and links it into its blockchain. It broadcasts the new hash to all full nodes, who confirm it against their respective replicas. Discrepancies are reported to the audit node. Transactions that arrive after the maximum block size is reached await the following 10 minute cycle. The coinbase transaction specifies as its unspent output, the particular Bitcoin address controlled by the reward agent.

The mint agent may perform merged mining on behalf of any compatible altcoin, e.g. a proof-of-stake version of Namecoin.

### 6.9. Reward Agent.

The reward agent distributes dividends daily to full nodes. By default, full nodes are compensated for their operating costs by a fixed proportion, and additionally receive a reward in proportion to their offered stake. Super-peer nodes receive a reward compensating for the infrastructure expenses they incur when provisioning the network backbone. These expenses may reasonably include media advertising, marketing support, and compelled contributions to various bitcoin non-profits, e.g. paying the salaries of operators monitoring the network, and developers enhancing Bitcoin. Reward allocation policies and other systemic policy decisions shall be periodically decided by humans elected into a certain paid-for non-profit organization by the stake-weighted votes of full nodes.

### 6.10. Primary Audit Agent.

The primary audit agent is a singleton having the responsibility of passive and active auditor. It receives reports of inconsistencies from any other node, e.g. some disagreement with consensus, and performs an investigation. If recovery is required, it messages the recovery agent concerning the nature of the recovery, e.g. revert the new block. The audit agent has the power to ban misbehaving nodes.

The audit agent daily polls each of the full nodes at random times, to ensure that they have the correct blockchain hash, and whether the entire blockchain is present, e.g. blockchain bytes at certain random offsets.

### 6.11. Secondary Audit Agents.

The secondary audit agents perform random checks that duplicate the intent of the primary audit agent and in certain cases offload audit tasks for parallelization.

### 6.12. Recovery Agent.

The recovery agent is a singleton having the responsibility of performing fault-recovery. Typical fault recovery in case of a defective mint would be the coordinated roll-back of the last committed block, releasing its transactions for input into a replacement block by a backup temporary mint.

The recovery node has the power to disable a faulty super-peer and replace it with a backup.

### 6.13. Network Operations Agent.

The network operations agent is a singleton having the responsibility of reporting node, super-peer, and network performance and integrity indicators, such as number of connected nodes, node churn rate, bandwidth consumed for various purposes, storage consumed, response time, outages, detected DDoS attacks, detected Sybil attacks, misbehaving nodes, alerts, etc. It provides a JSON API that enables a reference web site or any other web page to easily display these indicators as figures or charts in a variety of natural languages. It has command and control responsibilities for the common physical assets of the network. Humans are in the loop to a certain limited extent.

Possibly, the network operations agent migrates every few hours, following local traffic peaks between world financial capitals in successive time zones. For example, in order: Sydney, Tokyo, Hong Kong, Singapore, Moscow, Zurich, London, New York and San Francisco.

### 6.14. Software Provisioning Agent.

This agent is responsible for coordinating the deployment and reversion of software releases, in particular to super peer nodes.

## 7. System Test Plan.

A system test plan will be written in advance of coding the proposed changes. An orchestrated test harness, preferably using existing tools, will permit highly automated regression testing of a distributed set of Bitcoin software instances with a robust test suite. Early development work will focus on the tamper-evident logs and a query API added to the software that permits tracing a transaction through the network.

## 8. Conclusion.

The proposed reengineering is much more complex than Satoshi Nakamoto's Bitcoin. However, the complexity of the proposed system offers compelling benefits - mainly the avoidance of thermodynamic effort which is currently $490,103,250 annual rate, and growing an average 10x each year. Rewards are reallocated from proof-of-work, to the providers of decentralized network infrastructure, and to stakeholders as dividends. The paid-for enterprise network design is scalable to handle all the world's transactions.

Much of the complexity of the proposed system results from verifiable untrusted cooperation, allowing commonsense teamwork to rationalize the routing of transactions to specialized peers, and the synchronous maintenance of a replicated definite blockchain. As a result, customers and merchants receive sub-second transaction acknowledgement.

The proposed Bitcoin system is to be deployed in early 2016 as a hard fork of the blockchain within the Bitcoin network, following a year of public system testing. Launch is conditioned on wide acceptance among Bitcoin users, payment processors, developers, exchanges, hosted and customer-operated wallet providers, and the Bitcoin media.

**References.**